\documentclass[prc,twocolumn,showpacs,floatfix,superscriptaddress]{revtex4}
\usepackage{hyperref}
\usepackage{graphicx}
\usepackage{amssymb}
\usepackage{amsmath}
\usepackage{enumerate}
\usepackage{multirow}
\usepackage{hhline}
\usepackage{color}

\newcommand{\helium}[1][4]{{}^{#1}\mathrm{He}}
\newcommand{\beryllium}[1][8]{{}^{#1}\mathrm{Be}}
\newcommand{\symbBe}{\mathrm{Be}}
\newcommand{\carbon}[1][12]{{}^{#1}\mathrm{C}}
\newcommand{\oxygen}[1][16]{{}^{#1}\mathrm{O}}
\newcommand{\calcium}[1][40]{{}^{#1}\mathrm{Ca}}
\newcommand{\magnesium}[1][24]{{}^{#1}\mathrm{Mg}}

\begin{document}

\title{Monopole oscillations in light nuclei with a molecular
  dynamics approach}

\author{T.~Furuta}
\affiliation{
LPC Caen (CNRS-IN2P3/ENSICAEN et Universit\'e), F-14050 Caen, France
}
\author{K.~H.~O.~Hasnaoui}
\affiliation{
Department of Physics, Tohoku University, Sendai
980-8578, Japan
}
\author{F.~Gulminelli}
\affiliation{
LPC Caen (CNRS-IN2P3/ENSICAEN et Universit\'e), F-14050 Caen, France
}
\author{C.~Leclercq}
\affiliation{
LPC Caen (CNRS-IN2P3/ENSICAEN et Universit\'e), F-14050 Caen, France
}
\author{A.~Ono}
\affiliation{
Department of Physics, Tohoku University, Sendai
980-8578, Japan
}

\begin{abstract}
Collective monopole vibrations are studied in the framework of
an antisymmetrized version of molecular dynamics as a function of the
vibration amplitude. The giant monopole resonance energy in $\calcium$
is sensitive to the incompressibility of the effective interaction, in
good agreement with complete time-dependent Hartree-Fock calculations.
The collective response of $\carbon$, $\oxygen$, and $\magnesium$ is
also studied.  For these lighter nuclei that have an important
contribution of an $\alpha$-clustered component, different frequencies
are observed, corresponding to two different types of vibrations
associated with breathing and moving of the underlying clusters.
Possible connections with direct breakup into $alpha$ clusters at high
excitation energy are discussed.
\end{abstract}

\maketitle

\section{Introduction}
Collective vibrations of nuclei are interesting phenomena in quantum
many-body dynamics. The constituent nucleons coherently participate in
these vibrations and the motion can be intuitively understood in
classical terms as a global vibration of an excited liquid drop. In a
quantum mechanical sense, the dynamics is described by a coherent
superposition of many particle-hole excitations on top of the
ground-state configuration. These collective vibrations have been
experimentally observed for thirty years as broad resonances (giant
resonances) in the high-energy spectrum of nuclei \cite{pearson,
  shlomo93,young01a,young04}.  The existence of such resonances has
been reported for many different medium-heavy and heavy nuclei in the
isotopic table, and their characteristics have been successfully
reproduced in mean-field-based approaches \cite{blaizot}.

These theoretical studies have shown that collective modes are
privileged probes of the effective interaction, and as such they can
bear important information on nuclear matter properties. In
particular, the isoscalar giant monopole resonance or the breathing
mode of nuclei is connected to the incompressibility of nuclear
matter, even if this connection is far from being direct
\cite{shlomo04,colo04}.

Experimental data on the isoscalar monopole response of nuclei as
light as $\carbon$ or $\oxygen$
\cite{young98,young03,young01b,young09} show a fragmented strength
distribution, which is qualitatively understood as a global weakening
of the collectivity but is not quantitatively understood.  It is well
known that correlations play an important role in the structure of
these light nuclei and the application of standard Hartree-Fock (HF)
or random phase approximation (RPA) calculations to their collective
dynamics may be questionable. This is particularly true in nuclei
presenting clustered structures, a phenomenon that is expected both in
excited states and in the ground state of numerous light and
medium-heavy $N=Z$ nuclei \cite{ikeda,yamada04}.  In such situations,
one can expect that collective vibrations may imply collective motions
of the underlying clusters.  Indeed, it has already been reported in
the literature that collective strengths to low-lying states are
affected by the cluster degrees of freedom \cite{yamada08,kawabata07}.

The giant monopole resonance (GMR) corresponds to relatively small
amplitude of a collective monopole excitation. Another interesting
regime of nuclear dynamics is given by large-amplitude collective
motion with a significant compression of the system followed by
expansion.  At excitations of the order of 3 $A$MeV or higher, it
leads to the observable phenomenon of nuclear fragmentation
\cite{cugnon,friedman88,friedman90,chomaz04}.

Here the situation is qualitatively similar to the excitation of giant
resonances: When a heavy nucleus is compressed significantly, the
dynamics of compression and expansion appears to be governed by global
collective properties, namely, the nuclear matter equation of state,
and appears to be largely independent of the structure properties of
the excited nucleus \cite{isis,indra}.  Conversely, if the
ground state and/or excited states have pronounced clustered
structure, one may expect an interplay between these structures and
the collective dynamics and fragmentation pattern. A systematic study
of the fragmentation of light clustered nuclei has never been done,
but some indications exist as a preferential $\alpha$-particle
emission from $\alpha$-structured nuclei
\cite{INDRA08,radutaArX,kokalova06,oertzen08}.

In this respect, models able to describe both the small- and
large-amplitude collective regime are useful tools for understanding
collective nuclear dynamics.  Interesting studies have been performed
in the context of time-dependent Hartree-Fock (TDHF) calculations
\cite{chomaz87,giai88,reinhard06,naka05,naka07,stevenson07,avez,sim03,sim09}
and its semiclassical counterpart
\cite{smerzi,balbutsev99,balbutsev07,gaitanos10} and a cluster-based
time-dependent approach \cite{TDCM1,TDCM2,TDCM3}.

Antisymmetrized versions of molecular dynamics models, antisymmetrized
molecular dynamics (AMD) \cite{onoPRL,onoPTP} and fermionic molecular
dynamics (FMD) \cite{FMD}, are powerful approaches for describing both
nuclear structure and reactions.  They are microscopic models based on
nucleonic degrees of freedom. The wave function is modeled as a fully
antisymmetrized product of Gaussian wave packets.  The choice of the
nonorthogonal Gaussian wave packets is a restriction of general Slater
determinants, but it has an advantage of allowing spontaneous breaking
of all spatial symmetries. Because of that, these models are
particularly suitable to describe cluster and molecular states. After
projection over appropriate quantum numbers, they have been
successfully applied to describe the structure change from
shell-model-like structure to clustering structure
\cite{enyo99,enyo06,FMD}. They also have been applied to nuclear
reactions from fusion \cite{FMDreact} to multifragmentation reactions
\cite{onoPRL,onoPTP,onoReview}. For small-amplitude collective modes,
the isovector dipole oscillation has been studied with AMD
\cite{enyoGDR} but almost no application to the isoscalar monopole
modes has been performed to our knowledge.

In this paper, we use the simplest version of the model employing a
single Slater determinant. This approximation is not sufficient to
realistically describe nuclear spectroscopy, but it allows us to
understand qualitatively the effect of clustering on the collective
dynamics.  The use of a fully time-dependent approach allows us in
principle to study consistently the motion from the small-amplitude
RPA regime to the large-amplitude fragmentation regime.  In this first
application, two-body collisions and quantum branching, which are
necessary to explain experimental data of heavy-ion collisions around
Fermi energies \cite{onoPRL,onoPTP,onoAMD-V,onoReview}, are not
implemented, but their inclusion is straightforward.

The main difference between AMD and FMD concerns the width degree of
freedom. In FMD, the centroids and width of Gaussians are both complex
time-dependent variational parameters whereas, in AMD, the width is an
external real number common to all the Gaussians.  The isoscalar
monopole oscillation is supposed to correspond to a breathing mode, so
that the width degree of freedom may have a significant importance. By
comparing the results between AMD and FMD, we would also like to
clarify what degrees of freedom are important to properly describe
compression modes.

\section{Formalism of AMD and FMD}
\label{formalismAMDFMD}

In the AMD and FMD models, the optimal time dependence of the $N$-body
wave function is obtained through the time-dependent variational
principle
\begin{equation}
\begin{split}
\delta\int_{t_1}^{t_2}
\frac
{\langle\Psi(t)|i\hbar\frac{d}{dt}-\hat{H}|\Psi(t)\rangle}
{\langle\Psi(t)|\Psi(t)\rangle}
\,dt&=0\\
\delta|\Psi(t_1)\rangle=\delta|\Psi(t_2)\rangle&=0
\end{split}
\label{eq:TimeVariation}
\end{equation}
within the constraint that $|\Psi(t)\rangle$ is given by a Slater
determinant of the single-particle states $\psi$,
\begin{equation}
|\Psi\rangle=\frac{1}{\sqrt{A!}}\det_{ij}\left[\psi_i\;(j)\right],
\end{equation}
where $A$ is number of nucleons.

If no extra restriction on the single-particle wave functions
$\psi(\vec{r})$ is applied, Eq.~(\ref{eq:TimeVariation}) gives the
TDHF equation \cite{ringschuck}.  However, in the AMD and
FMD models, the single-particle wave functions $\psi(\vec{r})$ are
restricted to the Gaussian form:
\begin{equation}
\psi_i(\vec{r})=
\exp\left[-\frac{1}{2a_i}\left(\vec{r}-\vec{b}_i
\right)^2\right]\chi_{\alpha_i} 
\label{Gaussian}
\end{equation}
where $\vec{b}_i$ ($i=1,\dots,A$) are complex time-dependent
variational parameters.  In the FMD formulation, the wave packet width
$a_i$ is an additional complex time-dependent variational
parameter. In contrast, in the AMD model, all wave packets share the
same real constant width $a_i=1/2\nu=\text{const}$.  The spin-isospin
states $(\chi_{\alpha_i}= p\uparrow, p\downarrow, n\uparrow,
n\downarrow)$ give additional variational parameters, which are
essential to describe charge-exchange reactions and spin excitations.
In the present application to scalar-isoscalar collective modes, only
spin and isospin saturated systems will be considered. We will then
restrict ourselves to a simple variation in the configuration space
and the spin-isospin states will be treated as constants of motion.
The equations of motion for AMD and FMD as resulting from
Eq.~(\ref{eq:TimeVariation}) can be written as
\begin{equation}
i\hbar\sum_{j\tau}C_{i\sigma,j\tau}\frac{dq_{j\tau}}{dt}
=\frac{\partial \langle\hat{H}\rangle}{\partial q_{i\sigma}^\ast}
\end{equation}
where $\langle\hat{H}\rangle$ is the expectation value of the
Hamiltonian $\hat{H}$ taken with the wave function $\Psi$,
$C_{i\sigma,j\tau}$ appears because of the non-orthogonality of
Gaussian wave packets and is given by
\begin{equation}
C_{i\sigma,j\tau}=\frac{\partial^2} {\partial q_{i\sigma}^\ast\partial
  q_{j\tau}} \ln\langle\Psi|\Psi\rangle,
\end{equation}
and $q_{i\sigma}$ give the variational parameters for nucleon $i$ in
the model, namely, $q_{i\sigma=1,\dots,4}=a_i,b_{ix},b_{iy},b_{iz}$ in
FMD and $q_{i\sigma=1,\dots,3}=b_{ix},b_{iy},b_{iz}$ in AMD.  For
further details, see Refs.~\cite{onoReview,FMD}.

\section{Description of ground-state properties}
\label{groundstate}

The ground state is obtained by minimizing the energy expectation
value $\langle\hat{H}\rangle\equiv
{\langle\Psi|\hat{H}|\Psi\rangle}/{\langle\Psi|\Psi\rangle}$ with
respect to the variational parameters of the models.

The molecular dynamics ansatz Eq.~(\ref{Gaussian}) represents a
restriction of the full space for the single-particle wave
function. This means that a full HF calculation in three-dimensional
$\mathbf{r}$-space is a generalization of this model, though molecular
dynamics can be superior to HF if the latter is solved on a truncated
basis or by imposing the conservation of chosen spatial symmetries
\cite{boncheEV8}. As a consequence, AMD and FMD ground-state energies
have to be seen as upper limits of the HF ground-state value.

Table \ref{tb:GSprop} shows some ground-state properties, namely, the
ground-state energy $E_\text{g.s.}$ and the matter root-mean-square
radius $R_m$ obtained by AMD and FMD for several selected
nuclei. Results with SLy4 \cite{chabanat98}, Gogny \cite{gogny}, and
the old SIII effective interaction \cite{beiner75} are shown.  SLy4
and SIII are Skyrme interactions, namely, of $\delta$-type, with
symmetric nuclear matter incompressibilities of $K_\infty=230$ MeV and
$K_\infty=355$ MeV, respectively. Gogny is a finite-range interaction
with an incompressibility modulus of $K_\infty=228$ MeV.  Both Gogny
and SLy4 are realistic interactions that have been successfully
compared to different experimental data on nuclear masses and
collective modes. SIII also gives reasonable values for nuclear masses
along the stability valley, but it is known to exhibit a too high
incompressibility.  The results of a three-dimensional HF calculation
with SLy4 using the code of Ref.~\cite{boncheEV8} and the experimental
data from the Atomic Mass Evaluation \cite{audi} are also shown in the
table for comparison.
\begin{table}[thbp]
\begin{center}
 \begin{tabular}{|c|c|c|c|c|c|c|c|c|}
\hline 
\multicolumn{2}{|c|}{}  &
\multicolumn{3}{|c|}{AMD} & \multicolumn{2}{|c|}{FMD} & HF & Exp \\ 
\hhline{|~~|-|-|-|-|-|-|-|}
\multicolumn{2}{|c|}{} & SLy4 & SIII & Gogny & SLy4 & SIII & SLy4 & \\ 
\hline 
\multirow{2}{*}
{$\helium$} &
$E_{\text{g.s.}}$ & -26.0 & -25.6 & -28.9 &
-26.2 & -25.5 & -26.7 & -28.3 \\ 
\hhline{|~|-|-|-|-|-|-|-|-|}
&$R_m$ & 1.73 & 1.69 & 1.63 & 1.73 & 1.69 & 1.96
& \\ 
\hline 
\multirow{2}{*}
{$\carbon$} & 
$E_{\text{g.s.}}$ & -72.2
& -70.6 & -74.9 & -77.4 & -75.2 & -90.7 &
-92.2\\ 
\hhline{|~|-|-|-|-|-|-|-|-|} 
&$R_m$ &
2.57 & 2.53 & 2.52 & 2.57 & 2.51 & 2.44&\\ 
\hline
\multirow{2}{*} {$\oxygen$} & $E_{\text{g.s.}}$ & -121.5 &
-120.4 & -125.3 & -127.9 & -125.4 & -128.6 &
-127.6\\ 
\hhline{|~|-|-|-|-|-|-|-|-|} 
&$R_m$ &
2.63 & 2.59 & 2.58 & 2.63 & 2.60 & 2.67 & \\ 
\hline
\multirow{2}{*}{$\calcium$} & 
$E_{\text{g.s.}}$ & -330.6 &
-322.1 & -334.2 & -338.9 & -330.3 & -344.6 &
-342.1\\ 
\hhline{|~|-|-|-|-|-|-|-|-|} &
$R_m$ &
3.40 & 3.42 & 3.36 & 3.40 & 3.41 & 3.39 & \\ 
\hline
\end{tabular}
\end{center}
\caption{Ground-state properties [energy $E_{\text{g.s.}}$ (MeV) and
    root-mean-square radius $R_m$ (fm)] for $\helium$, $\carbon$,
  $\oxygen$ and $\calcium$ calculated by AMD and FMD with three
  different effective interactions (SLy4, SIII, and
  Gogny). Calculation results of the HF method with SLy4 using the
  code of Ref.~\cite{boncheEV8} and experimental data (EXP.) from the
  Atomic Mass Evaluation \cite{audi} are also shown in the table for
  comparison.}
\label{tb:GSprop}
\end{table}

\begin{table}[thbp]
\begin{center}
 \begin{tabular}{|c|c|c|c|}
\hline 
AMD $\nu_{\text{g.s.}}$ & SLy4 & SIII & Gogny \\ 
\hline 
$\helium$ & 0.19  & 0.20  & 0.21 \\ 
\hline 
$\carbon$ & 0.155 & 0.16  & 0.16 \\
\hline
$\oxygen$ & 0.155 & 0.16  & 0.16 \\
\hline
$\calcium$& 0.125  & 0.125 & 0.13 \\
\hline
\end{tabular}
\end{center}
\caption{ List of the inverse width optimal value
  $\nu_{\text{g.s.}}$ ($\text{fm}^{-2}$) for the AMD calculations.}
\label{tb:GSwidth}
\end{table}

In AMD calculations, the inverse width $\nu$ is an external parameter
optimized for each nucleus to ensure a minimum
$\langle\hat{H}\rangle$.  The optimal values $\nu_\text{g.s.}$ are
listed in Table \ref{tb:GSwidth}. The main difference between AMD and
FMD in this context, therefore, is that (i) AMD takes a common width
for all the Gaussians whereas FMD can take a different width for each
Gaussian and (ii) the imaginary part of the width is an extra
variational parameter for FMD whereas it is fixed to zero in AMD.  In
actual applications, point (i) is the most important since it helps in
describing neutron skins and neutrons halos
\cite{FMDst04,FMDst05,FMDst07}. However, the imaginary parts of the
variational parameters turn out to be very close to zero in FMD ground
states, even if they can play a role in the dynamical evolution.

As was already observed in previous more systematic studies
\cite{onoReview,FMD}, the overall experimental features of the
ground-state properties are reproduced by both AMD and FMD.  The
difference between the results of HF and those of AMD and FMD is not
significant for most of the cases, even though the state
$|\Psi\rangle$ in AMD and FMD does not cover the whole possible range
of Slater determinants. This result shows that the Gaussian ansatz has
a high degree of flexibility, at least as far as the shown observables
are concerned.

The only exception is given by the case of $\carbon$, which is
significantly underbound by both AMD and FMD.  This discrepancy is not
due to the limited variational space of the spatial wave functions,
but rather to the lack of the spin-orbit interaction, which has been
neglected in the present work.  In fact, if we switch off the
spin-orbit interaction in the HF calculation, we obtain
$E_{\text{g.s.}}=-74.8$ MeV, which is close to the FMD result.  This
value is slightly higher than the FMD result, which may be due to the
mirror symmetry in $xyz$ axes imposed on the single-particle wave
functions in the HF calculation \cite{boncheEV8}.  To properly
introduce the spin-orbit interaction into FMD (and AMD), the variation
of the spin wave function together with the configuration wave
function \cite{FMD}, as well as the angular momentum projection, would
be necessary \cite{enyo05,FMDst04,FMDst05,FMDst07}.  These features
are essential for a quantitative study of nuclear structure and
spectroscopy, but they are not expected to play an important role in
the study of the monopole oscillation.
 
Because of the restriction of the accessible configuration space in
molecular dynamics models, the following relation is expected among
the ground-state energies calculated by these models if the same
effective interaction is used:
\begin{equation}
E_\text{g.s.}(\text{HF})\leq E_\text{g.s.}(\text{FMD})\leq
E_\text{g.s.}(\text{AMD}).
\end{equation} 
However, this inequality is not systematically observed.  This is due
to the different treatment of the spurious zero-point kinetic energy
of the center-of-mass motion $T_{\text{zero}}$.  It is well known that
the center-of-mass motion gives rise to a spurious contribution in
HF-based models, owing to the breaking of translational
invariance. The first-order contribution can be subtracted by
replacing the kinetic part of the Hamiltonian $\hat{H}_\text{kin} =
\frac{1}{2m}\sum_{i=1}^A\hat{\vec{p}_i}^2$ with

\begin{equation}
\begin{split}
\hat{H}_\text{kin}
&-\frac{1}{2mA}\biggl(\sum_{i=1}^A\hat{\vec{p}}_i\biggr)^2\\
=&
\hat{H}_{\text{kin}}-\frac{1}{2mA} \biggl(
\sum_{i=1}^A\hat{\vec{p}_i}^2 +
\sum_{i\neq j=1}^A\hat{\vec{p}}_i\cdot\hat{\vec{p}}_j\biggr),
\end{split}
\end{equation}
where $m$ is the mass of the nucleon. The one-body term is readily
implemented by multiplying a factor $(1-1/A)$ to the kinetic energy.
The two-body term is numerically expensive, and it is neglected both
in the standard choice of HF \cite{boncheEV8} and in our FMD
calculations. Conversely, in AMD, the center-of-mass coordinate can be
separated from the intrinsic degrees of freedom and there is no
spurious coupling between them. Therefore, the contribution of the
spurious center-of-mass motion is readily calculated as
\begin{equation}
T_{\text{zero}}=\frac{1}{2mA}
\frac{
\langle \Psi_\text{AMD}|(\sum_{i=1}^A\hat{\vec{p}}_i)^2|
\Psi_\text{AMD}\rangle}
{\langle \Psi_\text{AMD}|\Psi_\text{AMD}\rangle}
=\frac{3\hbar^2 \nu}{2m}
\end{equation} 
and can be exactly subtracted from the AMD ground-state energy.  This
approximate treatment of the center-of-mass motion in HF and FMD can
be of some importance in the case of the light nuclei considered here.
When employing in the AMD calculation the same approximate treatment
of the center-of-mass motion that is currently used in FMD and HF, we
have verified that indeed $E_{\text{g.s.}}(\text{HF})\leq
E_{\text{g.s.}}(\text{FMD})\leq E_{\text{g.s.}}(\text{AMD})$ is always
found.

\section{Description of the isoscalar monopole collective motion}
\label{gmr}

Collective motions are easily identified in time-dependent approaches
as periodic oscillations of the associated collective variables.  For
$L=0$ excitations, an initial perturbation preserving radial symmetry
induces a collective vibration of the nucleus with specific
frequencies that can be extracted from a Fourier analysis of the
oscillation pattern.  The collective vibration of a spherical nucleus
in small amplitude is typically associated with a unique frequency,
the giant monopole resonance, whose energy is directly related to the
second derivative of the energy functional with respect to the
root-mean-square radius, that is, to the nucleus incompressibility
coefficient \cite{blaizot}.

The advantage of explicit time-dependent approaches compared to the
more standard RPA analysis is that they allow multiple excitations of
giant resonance and coupling to more complex high-energy channels
\cite{balbutsev99,balbutsev07}.  Since at large amplitude the monopole
oscillation evolves toward radial expansion followed by nuclear
multifragmentation, we may expect a correlation between the
oscillation frequencies and the expansion speed which may affect the
fragmentation pattern \cite{chikazumi}.  In the standard
time-dependent treatment of collective excitations
\cite{stringari79,chomaz87,pacheco88}, the initial state
$|\Psi(t=0)\rangle$ is prepared by applying the generator of the
collective excitation under study, $\hat{U}_0(k)=\exp(ik\hat{Q}_n)$,
on the ground-state wave function $\langle
\vec{r}_1,\dots,\vec{r}_A|0\rangle$ , where $\hat{Q}_0=\hat{r}^2$ for
the isoscalar monopole.  In this case the Fourier transform of the
temporal variation of the collective response $\Delta\sqrt{\langle
  \hat{r}^2 \rangle}=\sqrt{\langle \hat{r}^2 \rangle(t)} - \sqrt{
  \langle \hat{r}^2 \rangle (t=0)}$,
\begin{equation}
\text{FT} [r](\omega)=
\int_{-\infty}^{\infty} dt\Delta \sqrt{ \langle \hat{r}^2 \rangle(t) }
e^{-i\omega t}
\end{equation}
in the small-amplitude limit ($k\to 0$) is directly proportional to
the strength function
\begin{equation}
\Im \left ( \text{FT} [r](\omega)\right ) \propto S(\omega)=
\sum_n  |\langle 0|\hat{r}^2|n \rangle|^2 
\delta \left ( \hbar\omega- E_n \right ), 
\label{eq:strength}
\end{equation}
meaning that the excitation probability of the different frequencies
is a direct measure of the transition probability to the excited
states $|n\rangle$ of energy $E_n$, induced by the excitation operator
$\hat{Q}_0$.

This treatment cannot be applied to the AMD model. Indeed the
application of the monopole operator $\hat{U}_0(k)$ to the
single-particle wave function, Eq.~(\ref{Gaussian}), produces a new
Gaussian state with centroid $\vec{b}'_i$ and width $a'_i$:
\begin{equation}
\vec{b}'_i=\frac{1+2ika^*_i}{1+4k^2|a_i|^2}\vec{b}_i \; \; ; \; \; a'_i
=\frac{1+2ika^*_i}{1+4k^2|a_i|^2}a_i 
\label{eq:fmd_init}
\end{equation}
which, because of the nonzero imaginary part of the width, does not
belong to the AMD variational space.  This simple observation already
suggests that the dynamical evolution of the width may play a
significant role in the breathing mode.

However, we can still study compressional collective modes in AMD by
choosing the initial state as a common translation in the radial
direction of the Gaussian centroids with respect to their ground-state
value,
\begin{equation}
\begin{split}
\Re\left(\vec{b}_i(t=0)\right)&=(1+\lambda)\Re(\vec{b}_i^{0})\\
\Im\left(\vec{b}_i(t=0)\right)&=\Im(\vec{b}_i^{0}).
\end{split}
\label{eq:amd_init}
\end{equation}

This initial state, as any other modifying the root-mean-square radius
of the nucleus, will excite the same monopole eigenmode as the
application of the monopole boost $\hat{U}_0(k)$ \cite{sim03}.  This
is however not true as far as the spectral function $\text{FT}
[r](\omega)$ is concerned. Indeed in the case of the AMD model where
the Gaussian centroids are the only variational parameters, this
initial state corresponds to minimizing the total energy in the
presence of a perturbative external field $-\lambda r^2$ for some
chosen values of $\lambda$.  It is possible to show
\cite{stringari79,chomaz87} that including an external harmonic field
$V_{ext}=\lambda \hat{r}^2 \theta\left ( -t \right )$ in the TDHF
dynamics is indeed an alternative way to prepare the initial monopole
state. Such an initial condition produces in the linear response
regime a spectral function different from Eq.\ (\ref{eq:strength}),
where the probability of the different collective states is weighted
by the inverse of the state energy.

For this reason we will concentrate on the energy of the collective
states, which appears to be independent of the chosen initial
condition, rather than on the detailed shape of the strength
function. Another argument for that is given by the fact that a
realistic description of the strength function and the associated
energy weighted sum rule (EWSR) requires a proper treatment of the
damping of collective motion, which in turn demands the inclusion of
higher order correlations in the form of collisions and branching
\cite{onoReview}.  Finally, since the equations of motion of AMD and
FMD are obtained by restricting the TDHF single-particle wave
functions to Gaussians, the strength function obtained by AMD or FMD
will approximately satisfy the corresponding RPA sum rule only if the
restriction on the single-particle wave functions is not critical.
All these difficulties are much less critical as far as the
frequencies are concerned. For all the cases presented in this paper,
the time evolution of the root-mean-square nuclear radius presents
only a few well-distinct frequencies, meaning that the collective
energies can always be unambiguously recognized.

\begin{figure}[htbp]
\begin{center}
\includegraphics[width=\columnwidth]{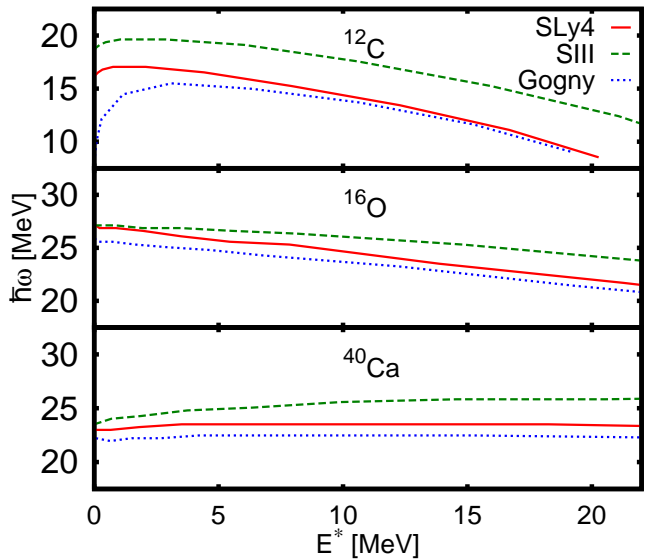}
\end{center}
\caption{(Color online) AMD prediction for the frequency $\hbar\omega$
  of the monopole oscillation as a function of the oscillation energy
  for $\carbon$, $\oxygen$, and $\calcium$ with the SLy4, SIII, and
  Gogny interactions. }
\label{fig:AMD3force}
\end{figure}

In AMD calculations for monopole oscillations of $\carbon$, $\oxygen$,
and $\calcium$, the spectral function $\text{FT} [r](\omega)$ shows
only a single peak and thus a single frequency in the oscillation. The
peak frequency $\hbar\omega$ as a function of oscillation excitation
energy, which is directly related to the oscillation amplitude, with
the three different interactions (SLy4, SIII and Gogny) are shown in
Fig.~\ref{fig:AMD3force}. A global independence of the peak frequency
on the excitation energy is observed in the case of $\calcium$,
whereas the peak frequency gradually decreases for $\carbon$ and
$\oxygen$ at high excitation energies (>$5$ MeV).  For low excitation
energy (small-amplitude) oscillations (<$5$ MeV), the peak frequency
varies abruptly as a function of excitation energy for the $\carbon$
case if the Gogny interaction is used.  A clear, even if much less
spectacular, frequency change for small amplitude oscillations is also
observed for the other effective interactions and the other nuclei.
This behavior will be discussed in detail in the next section.
However, this behavior at small amplitudes is not likely to strongly
affect the energy of one-phonon states when the collective motion is
quantized.  The physical frequency that is relevant to the breathing
mode is rather the one at the excitation energy of the mode itself,
since the percentage of the one-phonon state is maximized at this
energy.

For the case of $\calcium$, different theoretical studies exist on the
GMR in this nucleus \cite{avez,colo07}, which can thus be considered
as a benchmark for our calculations.  The ground state of $\calcium$
is obtained in AMD as a structureless spherical distribution of
neutron and proton densities.  The resulting monopole peak is found at
an energy $\hbar\omega$ in reasonably good agreement with TDHF
\cite{avez} ($\hbar\omega=22.1$ MeV) and RPA \cite{colo07}
($\hbar\omega=21.5$ MeV) calculations using the same SLy4 effective
interaction.

The SIII interaction is employed here for the purpose of exploring the
sensitivity of the monopole response to the nuclear matter properties.
As expected, as the incompressibility $K_\infty$ increases from 230
MeV (SLy4) to 355 MeV (SIII), all the peak frequencies increase,
showing that indeed collective compression and expansion can be
addressed in molecular-dynamics-based formalisms.

The peaks obtained by AMD turn out to be without a physical width
($\delta$-function-like). In principle the energy of the modes is
above the threshold of particle emission and the motion should be
damped by the coupling to the continuum
\cite{chomaz87,reinhard06,avez}.  This is however not observed in the
simulation, showing that the AMD wave function ansatz is not flexible
enough to properly describe particle evaporation.  The inclusion of
higher order correlations is also required for a proper treatment of
the damping of collective motions.  In future studies it would be
interesting to investigate whether the AMD models are able to
reproduce the width of the collective resonances by introducing the
two-body collisions and quantum branching \cite{onoReview}.

\section {Different modes and the role of the width}
\label{effect_width}

In AMD calculations, the inverse width parameter $\nu$ is an external
parameter and the calculations shown in Sec.~\ref{gmr} were performed
using the optimal value $\nu_{\text{g.s.}}$ (table \ref{tb:GSwidth}),
namely, the one that gives a minimum ground-state energy.  However,
there is no strong reason to adopt $\nu_{\text{g.s.}}$ for the
oscillation calculation if the collective excited states have
considerably different structure from the ground state. For example,
if the collective excited states have a strong clustered structure,
the optimal choice of $\nu$ for individual clusters may be more
suitable than $\nu_{\text{g.s.}}$.  Furthermore, in some applications
of AMD such as to nuclear matter and heavy-ion reactions, it is not
practical to adjust $\nu$ for individual nuclei that appear in the
system.  For these reasons, it is important to understand how the
characters of monopole oscillation depend or do not depend on the
choice of $\nu$, which we will study in this section.  We take the
$\carbon$ calculation as a reference in the following, but
qualitatively similar results are observed for $\oxygen$ and
$\magnesium$.
  
\begin{figure}[htbp]
\begin{center}
\includegraphics[width=\columnwidth]{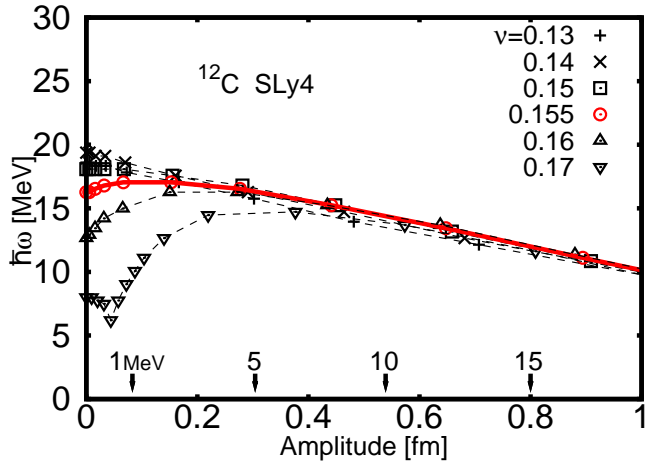}
\end{center}
\caption{(Color online) AMD prediction for the peak position
  $\hbar\omega$ of the monopole oscillation for $\carbon$ with the
  SLy4 interaction as a function of the oscillation amplitude for
  different values of the inverse width parameter $\nu$
  ($\text{fm}^{-2}$). The excitation energies corresponding to
  oscillation amplitudes in the case of the optimal width
  $\nu=0.155\;\text{fm}^{-2}$ (full line) are indicated by arrows. }
\label{fig:C12_HWvsR}
\end{figure}

Figure \ref{fig:C12_HWvsR} shows the position of the peak frequency in
the spectral function $\text{FT}[r](\omega)$ of $\carbon$ by AMD with
SLy4 as a function of oscillation amplitude for different inverse
width parameters $\nu$. The full line shows the result of the ground
state optimal width ($\nu_{\text{g.s.}}=0.155 \;\text{fm}^{-2}$). The
excitation energies corresponding to oscillation amplitudes for this
line are also indicated in the figure.

In the region of the amplitude greater than 0.4 fm, different choices
of $\nu$ give almost identical oscillation frequencies, and in
particular the prediction of GMR energy is almost independent of the
choice of $\nu$.  The frequency gradually decreases as the amplitude
increases. This behavior can be physically understood, since above the
breakup threshold the expansion can be followed by a breakup of the
system and not by a recompression, corresponding to an oscillation
period that tends to diverge (that is, a frequency that tends to
zero).

In small-amplitude oscillations ($<$0.4 fm), the oscillation frequency
varies abruptly as a function of the amplitude for some choices of
$\nu$.  A very peculiar behavior is observed in particular for the
case $\nu=0.17\;\text{fm}^{-2}$ in the small-amplitude regime.  With
this choice of the width parameter the monopole frequency shows a
well-pronounced minimum at a specific value of the oscillation
amplitude (0.04 fm).  In the following we will show that this is due
to the presence of two possible distinct oscillation regimes in the
AMD dynamics by investigating the motion of wave packet centroids.

\begin{figure}[htbp]
\begin{center}
\includegraphics[width=\columnwidth]{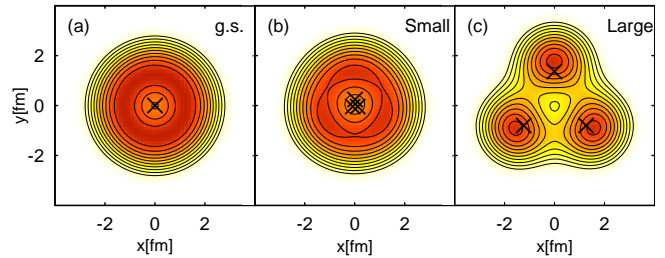}
\end{center}
\caption{(Color online) AMD density distributions of $\carbon$ with
  the SLy4 interaction projected on the plane perpendicular to the
  symmetry axis ($x$-$y$ plane) for (a) the ground state and the
  maximum amplitude stage of the monopole vibration at (b)
  $E^\ast=0.1$ MeV and at (c) $E^\ast=5$ MeV. The cross marks in the
  figures indicate the parameter positions of the effective
  ``$\alpha$''s (See text).}
\label{fig:densityC12}
\end{figure}

The motion of Gaussian centroids gives some useful information on the
time evolution of the system, though the centroids do not correspond
to the physical positions of nucleons owing to antisymmetrization.  In
the AMD dynamics of $\carbon$, three ensembles of four Gaussians with
different spin-isospin ($p\uparrow$, $p\downarrow$, $n\uparrow$, and
$n\downarrow$) take almost the same time-dependent value for their
$\vec{b}$ parameter.  The global $\carbon$ configuration shows
therefore a ``3-$\alpha$-triangular'' structure during the time
evolution. The density distribution does not show spatial clustered
structure at very small amplitude [Fig.~\ref{fig:densityC12}(b)]
because of the large overlap among the ``$\alpha$''s and the
antisymmetrization effect among the nucleons, but these structures are
clearly visible for larger amplitude vibrations
[Fig.~\ref{fig:densityC12}(c)].  We will speak of effective
``$\alpha$''s in the following, where the word ``effective'' means
that the associated four nucleons do not form three independent
$\alpha$ particles because of the global antisymmetrization of the
wave function.

\begin{figure}[htbp]
\begin{center}
\includegraphics[width=\columnwidth]{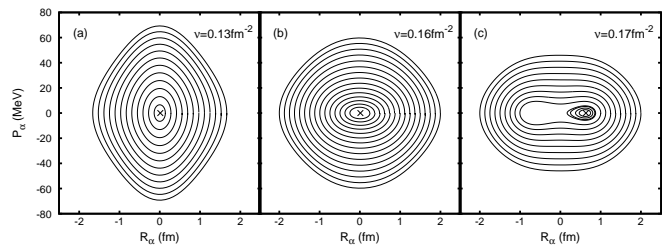}
\end{center}
\caption{Phase-space trajectory of the AMD monopole oscillation of
  $\carbon$ from the evolution of the variational parameters of one
  ``$\alpha$'' (see text) for (a) $\nu=0.13\;\text{fm}^{-2}$, (b) 0.16
  $\text{fm}^{-2}$, and (c) 0.17 $\text{fm}^{-2}$ with SLy4.  Each
  contour corresponds to a different initial expansion, the largest
  circle being associated with the largest amplitude oscillation. The
  global oscillation pattern of $\carbon$ can be deduced from symmetry
  considerations (see text). }
\label{fig:C12_traj}
\end{figure} 

\begin{figure}[htbp]
\begin{center}
\includegraphics[width=0.7\columnwidth]{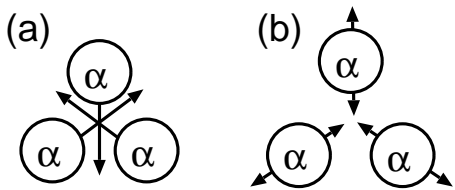}
\end{center}
\caption{Schematic representation of the two possible oscillation
  regimes observed in AMD calculations varying the Gaussian width
  and/or the motion amplitude.}
\label{fig:cartoon}
\end{figure}

The active degrees of freedom in this motion are the variational
parameters of one of these effective ``$\alpha$''s in $\carbon$,
namely, the Gaussian centroid parameters
\begin{equation}
\begin{split}
\vec{R}_\alpha &= 
\frac{1}{4}\langle\Phi_\alpha|\hat{\vec{r}}|\Phi_\alpha\rangle 
=\frac{1}{4}\sum_{i=1}^4\Re \left (\vec{b}_i\right )\\
\vec{P}_\alpha &=
\langle\Phi_\alpha|\hat{\vec{p}}|\Phi_\alpha\rangle =
2\hbar\nu\sum_{i=1}^4\Im \left (\vec{b}_i \right );
\end{split}
\end{equation}
the other two effective ``$\alpha$''s move in phase to keep the
triangular symmetry owing to the symmetry of the
excitation. $\vec{R}_\alpha={R}_\alpha\vec{u}_r$ and
$\vec{P}_\alpha={P}_\alpha\vec{u}_r$ oscillate along the radial
direction $\vec{u}_r=\vec{R}_\alpha(t=0)/|\vec{R}_\alpha(t=0)|$.  The
trajectories of $R_\alpha$ and $P_\alpha$ for three different choices
of the inverse width are shown in Fig.~\ref{fig:C12_traj}.  The
ground-state $(R_\alpha,P_\alpha)$ values are indicated by the crosses
in Fig.~\ref{fig:C12_traj}.  In the case of a wide Gaussian width
$\nu=0.13\;\text{fm}^{-2}$ [Fig.~\ref{fig:C12_traj}(a)] the position
of the effective ``$\alpha$''s in the ground state coincides with the
center of mass of the nucleus. The trajectories circulate in phase
space around $(R,P)=(0,0)$ and the other two effective ``$\alpha$''s
move in phase so that this motion corresponds to the crossing of
``3-$\alpha$'' as schematically shown by the cartoon of
Fig.~\ref{fig:cartoon}(a).
For narrower Gaussian width $\nu>0.16 \;\text{fm}^{-2}$, the ground
state $R_\alpha$ becomes finite, indicating an explicit ``$\alpha$''
structure even in the ground state. In this case the small-amplitude
oscillations [Fig.~\ref{fig:C12_traj}(c)] correspond to vibrations
around this finite value, leading to a trajectory confined to the
positive $R_\alpha$-coordinate space. This motion can be visualized as
three effective ``$\alpha$''s oscillating around their ground state
values as schematically indicated by Fig.~\ref{fig:cartoon}(b).
Increasing the excitation energy, and therefore increasing the
amplitude of the motion, we find trajectories circulating around
$(R_\alpha,P_\alpha)=(0,0)$ as in the case of $\nu=0.13
\;\text{fm}^{-2}$. The frequency minimum for $\nu=0.17
\;\text{fm}^{-2}$ in Fig.~\ref{fig:C12_HWvsR} is observed at the
transition between these different oscillation regimes.  The
small-amplitude oscillations observed with $\nu=0.16 \;\text{fm}^{-2}$
[Fig.~\ref{fig:C12_traj}(b)] are located at the transition between the
two motions, which explains the strong $\nu$ dependence of the peak
frequency for the small amplitude oscillations in
Fig.~\ref{fig:C12_HWvsR}.

As the time-dependent state $|\Psi_{\text{AMD}}[q(t)]\rangle$ violates
the symmetries, a more desirable approach is to apply the
time-dependent variational principle to the state
$P(0^+)|\Psi_{\text{AMD}}[q]\rangle$ after the parity and angular
momentum projection.  For the oscillations with two regimes as in
Fig.~\ref{fig:C12_traj}(c), the excitation from the ground state,
which has a finite ``$\alpha$'' clustering, should couple to the
excitation from the parity-reflected ground state.  The lack of such a
coupling in the present calculation may be a reason of the low
oscillation frequency around the transitional situation between
Figs.~\ref{fig:cartoon}(a) and \ref{fig:cartoon}(b). However, if the
oscillation amplitude is large (namely, if the ``deformation'' of the
intrinsic state $|\Psi_{\text{AMD}}[q(t)]\rangle$ is large enough), we
can expect that the effect of the parity and angular momentum
projection is not so important.  It should also be noted that a kind
of many-particle many-hole excitations (namely, cluster excitation) is
included in the one-phonon state of the oscillation if it is extracted
from the time-dependent many-body state
$|\Psi_{\text{AMD}}[q(t)]\rangle$ or
$P(0^+)|\Psi_{\text{AMD}}[q(t)]\rangle$ around the GMR amplitude.

\begin{figure}[htbp]
\begin{center}
\includegraphics[width=\columnwidth]{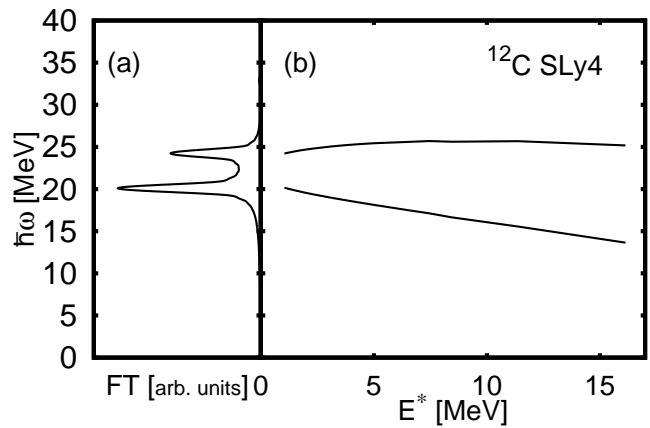}
\end{center}
\caption{(a) The spectral function $\text{FT} [r](\omega)$ in
  arbitrary units of the FMD monopole oscillation in small amplitude
  for $\carbon$ with the SLy4 interaction, obtained by the same
  initial state as in the AMD calculation. (b) The peak position
  $\hbar\omega$ as a function of the oscillation amplitude.}
\label{fig:FMD_C12}
\end{figure}

From these discussions, it is very interesting to see the results of
FMD, which explores dynamically the different values of wave packet
widths $\{a\}$ as well as the centroids $\{\vec{b}\}$.  Figure
\ref{fig:FMD_C12} shows the position of the peak frequency
$\hbar\omega$ in the spectral function $\text{FT} [r](\omega)$ of
$\carbon$ by FMD as a function of oscillation energy, where the
initial state is prepared by Eq.\ (\ref{eq:amd_init}). The spectral
function in small amplitude is shown on the left side of this figure.
The widths of the peaks are only a numerical artifact coming from the
time filter \cite{reinhard06}, which is applied to compensate for the
finite integration time in the Fourier transform.  In the case of FMD,
there are two peaks in the spectral function of $\carbon$ and the two
lines in Fig.~\ref{fig:FMD_C12}(b) correspond to the frequencies of
these peaks. The behavior of the lower frequency follows closely the
behavior of the one observed in AMD in the excitation energy regime
where no sensitivity to the width was observed. However, a smooth
behavior is also obtained at the lowest amplitude, showing that indeed
two types of vibrations could be physically present.

For comparison, the experimental monopole strength distribution
\cite{young03} shows a fragmented structure composed of at least four
different peaks between 13 and 45 MeV of excitation energy exhausting
around 27\% of the EWSR. The average energy in this excitation energy
range is $m_1/m_0=21.9$ MeV with an root-mean-square width of 4.8 MeV,
in qualitative but not quantitative agreement with our results. The
experimental data also show a low-energy peak that is associated with
the excitation of the Hoyle state. No strength at these low energies
is observed in our calculation, consistent with the fact that effects
beyond mean field have been shown to be necessary to correctly
reproduce the Hoyle state \cite{yamada04,enyo06}.
 
To understand the physical meaning of the two peaks observed in the
calculation, let us investigate the dynamics of variational parameters
of $\carbon$ in the model.  Similar to the case of AMD, three
ensembles of four Gaussians with different spin-isospin ($p\uparrow$,
$p\downarrow$, $n\uparrow$, and $n\downarrow$) take almost the same
time-dependent values for their $\vec{b}$ parameters and constitute a
``3-$\alpha$-triangular'' structure. The $a$ parameters associated
with each of the three ``$\alpha$''s oscillate in the same way.
Therefore, the active degrees of freedom in the FMD motion are given
by the centroid of the ``$\alpha$''s like in AMD,
\begin{align}
\vec{R}_\alpha &=
\frac{1}{4}\langle\Phi_\alpha|\hat{\vec{r}}|\Phi_\alpha\rangle =
\frac{1}{4}\sum_{i=1}^4\left\{\Re \left (\vec{b}_i\right ) + \frac{\Im
  \left (a_i\right )\Im \left( \vec{b}_i\right)}{\Re \left(a_i\right
  )} \right\} \nonumber\\ 
\vec{P}_\alpha &=
\langle\Phi_\alpha|\hat{\vec{p}}|\Phi_\alpha\rangle = \hbar
\sum_{i=1}^4\frac{\Im \left (\vec{b}_i\right )}{\Re \left (a_i\right
  )}
\end{align}
together with the width of the effective ``$\alpha$''s,
\begin{equation}
\begin{split}
\sigma_{R} &=
\langle\Phi_\alpha|\hat{r}^2|\Phi_\alpha\rangle =
\sum_{i=1}^4\frac{3|a_i|^2}{2\Re \left (a_i\right )} \\
\sigma_{P} &=
\langle\Phi_\alpha|\hat{p}^2|\Phi_\alpha\rangle =
\sum_{i=1}^4\frac{3\hbar^2}{2\Re \left (a_i\right )}.
\end{split}
\end{equation}
 
\begin{figure}[htbp]
\begin{center}
\includegraphics[width=\columnwidth]{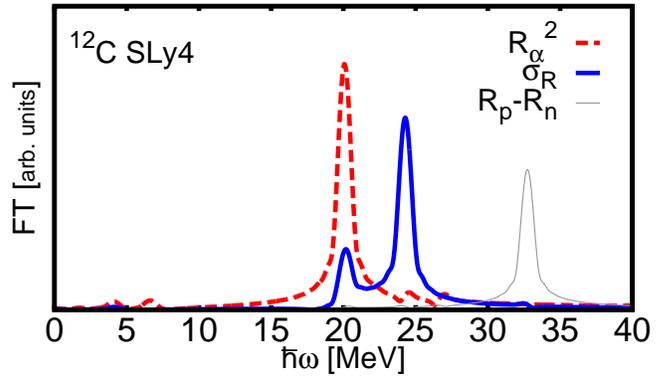}
\end{center}
\caption{(Color online) Fourier transforms in arbitrary units of FMD
  monopole oscillations ($\carbon$ with the SLy4 interaction)
  associated with the square of the position of the effective
  ``$\alpha$'' from the $\carbon$ center of mass $R_\alpha^2(t)$
  (dashed line); the coordinate width of the effective ``$\alpha$''
  $\sigma_{R}(t)$ (thick line); the differential time evolution of the
  proton and neutron for square radius
  $(R_\text{proton}-R_\text{neutron})$ (thin line).}
\label{fig:FMD_C12param}
\end{figure}

Figure \ref{fig:FMD_C12param} shows the oscillation frequencies
associated with these different observables.  We can see that the
first peak present in the monopole response (Fig.~\ref{fig:FMD_C12})
is associated with the time variation of the centroid of ``$\alpha$''
$R_\alpha$.  This physically corresponds to an oscillation of the
positions of three ``$\alpha$'' clusters around the center of the
nucleus [see Fig.~\ref{fig:C12_traj}(a) and
  Fig.~\ref{fig:cartoon}(a)].  In the absence of the width degree of
freedom, this is the only collective monopole motion accessible to the
system, and this is why the AMD response contains a single peak.  The
prediction of AMD and FMD concerning this peak is in good agreement
because it comes out that $\Im(a)\sim0$ during the whole dynamical
evolution.

The second peak at higher energy observed in the monopole response
(Fig.~\ref{fig:FMD_C12}) is associated with the time variation of the
widths.  It is physically associated with the coherent vibration of
the constituting ``$\alpha$'' clusters, as shown by the fact that this
frequency appears in the Fourier transform $\sigma_{R}(t)$. This mode
is very similar to the standard breathing mode of medium-heavy and
heavy nuclei, where all nucleons vibrate coherently.
 
In conclusion, there are two possible isoscalar monopole oscillations
in the FMD dynamics of an excited $\carbon$ nucleus: (1) cluster
moving and (2) cluster breathing.  This result is not unique for
$\carbon$, but it is also found for $\oxygen$. The cluster-originated
vibration modes are systematically found for all light nuclei that in
the framework of the AMD and FMD model exhibit a pronounced cluster
component. Vibration (1) finds its natural prolongation at high
amplitude in the phenomenon of direct cluster breakup, which is
systematically observed in our calculation at very high excitation
energy out of the linear response regime.

\begin{table}[thbp]
\begin{center}
\begin{tabular}{|c|p{1cm}|p{1cm}|p{1.8cm}|p{1.8cm}|}
\hline
\multirow{2}{*} {$\hbar\omega$ (MeV)} &
\multicolumn{2}{|c|}{initial state (\ref{eq:fmd_init})} &
\multicolumn{2}{|c|}{initial state (\ref{eq:amd_init})} \\
\hhline{|~|-|-|-|-|}
& SLy4 & SIII & SLy4 & SIII\\ 
\hline $^{12}$C & 24.2 & 30.4 & 20.2 \& 24.2 & 20.8 \& 30.7\\
\hline $^{16}$O & 26.1 & 32.4 & 26.2 \& 26.8 & 26.8 \& 32.8\\
\hline $^{40}$Ca & 22.1 & 27.6  & 20.9 & 27.5\\
\hline
\end{tabular}
\end{center}
\caption{Energy of the collective resonances $\hbar\omega$ recognized
  as peaks of the spectral function $\text{FT} [r](\omega)$ obtained
  by FMD for different selected light nuclei ($\carbon$, $\oxygen$,
  and $\calcium$) and effective interactions (SLy4, SIII, and Gogny)
  in the low-amplitude limit with two different initial states.}
\label{tb:GMR}
\end{table}

\begin{figure}[htbp]
\begin{center}
\includegraphics[width=\columnwidth]{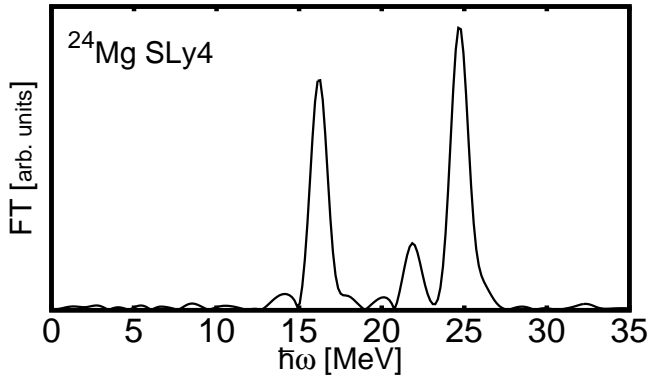}
\end{center}
\caption{The spectral function $\text{FT}[r](\omega)$ in arbitrary
  units of the FMD monopole oscillation in small amplitude for
  $\magnesium$ with the SLy4 interaction.}
\label{fig:Mg24FT}
\end{figure}

Table \ref{tb:GMR} shows the summary of the peak frequencies
$\hbar\omega$ in the spectral function $\text{FT} [r](\omega)$
obtained by FMD for different selected nuclei and effective
interactions in the small-amplitude limit with different initial
excitations. We can see that both the cluster moving and cluster
breathing modes are excited when the initial state is prepared by a
radial translation of the wave packets' centroids
[Eq.~(\ref{eq:amd_init})], whereas only the cluster breathing mode is
observed if the standard monopole boost [Eq.~(\ref{eq:fmd_init})] is
applied to the ground state FMD wave function in the small-amplitude
limit.  This suggests that the cluster moving mode in $\carbon$ may be
due to a coupling between the initial excitation
[Eq.~(\ref{eq:amd_init})] applied on a deformed ground state and the
monopole mode built on top of it, owing to the nonlinearity of the FMD
dynamics at finite amplitude. A similar effect was already reported in
TDHF calculations in Ref.~\cite{sim03}.  Moreover, in the case of
$\carbon$, the experimental threshold energy for 3-$\alpha$ breakup is
given by
$E^\ast=3E_{\text{g.s.}}(\alpha)-E_{\text{g.s.}}(\carbon)=7.3$ MeV
which is below the energy associated with the collective frequency of
the cluster moving mode of AMD and FMD, even considering the Coulomb
barrier, which is expected to be around 2 MeV \cite{yamada04}. The
observed vibration appears then more as a virtual diffusion state than
as a resonant state. This is however not the case for all cluster
moving modes. As an example we show in Fig.~\ref{fig:Mg24FT} the
Fourier transform $\text{FT}[r](\omega)$ of the FMD oscillation in
small amplitude for $\magnesium$.

The ground state of $\magnesium$ is found to be a deformed prolate
shape and it is natural to find a fragmented structure in the spectrum
function of such a nucleus since the monopole oscillation is strongly
coupled to the quadrupole oscillation. However, the analysis of the
variational parameters $\{a,\vec{b}\}$ reveals that once again the
peaks are related to different motions of the underlying
quasi-clusters.  Specifically, the origin of the three major peaks
($\hbar\omega=16$, 22, and 25 MeV) are seen to correspond with
3``$\symbBe$'', 2``$\symbBe$'' + 2``$\alpha$'', and 6``$\alpha$'',
respectively, where ``$\symbBe$'' denotes the effective cluster for
$\beryllium$ as ``$\alpha$'' for $\helium$.  These peak frequencies
are below the experimental threshold energy for 6-$\alpha$ breakup,
$E^\ast=6E_{\text{g.s.}}(\alpha)-E_{\text{g.s.}}(\magnesium)=28.5$ MeV
and therefore these resonances may physically exist, although the
implementation of the spin-orbit interaction and the projection over
angular momentum and parity before the variational procedure will be
necessary for a further quantitative discussion.

For comparison, the experimental monopole strength distribution
\cite{young09} of this nucleus shows a very broad structure peaked at
around 21 MeV with a root-mean-square width of around $5$-$7$ MeV,
exhausting around 80\% of the EWSR. Here again, only qualitative
conclusions can be advanced.  If the complicated structure of the
spectral functions that are systematically observed in light nuclei
\cite{young98,young03,young01b,young09} are certainly an indication of
a global loss of collectivity, our study suggests that a multiple-peak
monopole spectrum can also be an indication of clustered states. A
prolongation of this work requires the inclusion of collisions and
branching in the molecular dynamics \cite{onoReview}, which would
allow a quantitative prediction of the strength function as a possible
observable to evidence cluster structures.

Other collective modes can also be investigated with our approach by
choosing an appropriate initial state and analyzing the associated
observable. In fact, in the oscillation calculation of $\carbon$
discussed here, when adopting the initial state (\ref{eq:amd_init}),
we also find another peak around 33 MeV by looking at the Fourier
transform of the width variable of a proton wave packet.  This extra
peak is associated with the isovector monopole mode, which is clearly
seen in the Fourier transform of the differential time evolution of
the proton and neutron root-mean-square radius
$(R_\text{proton}-R_\text{neutron})$ (the thin line in
Fig.~\ref{fig:FMD_C12param}).  Indeed the chosen initial state, where
all the wave packets have the same width, contains some isovector
excitation in the FMD model because protons and neutrons have slightly
different widths in the ground state owing to the Coulomb force.

\section{Summary}
In this paper we have studied collective isoscalar
monopole oscillations for some selected light nuclei in the framework
of the AMD and FMD time-dependent approaches.

The study of GMR in $\calcium$ reveals that for medium-sized nuclei
molecular dynamics models are in good agreement with TDHF
calculations. A well-defined collective breathing mode is observed at
a frequency that depends on the incompressibility modulus of the
underlying effective interaction.
 
Clear collective modes are observed also for nuclei as light as
$\carbon$, but the collective modes in light nuclei appear dominated
by quasimolecular cluster structures. We have investigated in great
detail the motion of $\carbon$ as an example. The $\carbon$ isoscalar
monopole dynamics leads to a spectral function containing two distinct
frequencies associated, respectively, with an oscillation of the
effective three ``$\alpha$'' constituents around the nucleus center of
mass and with a coherent vibration of the underlying clusters. This
latter frequency sensitively increases with increasing
incompressibility of the associated equation of state, but does not
follow the empirical $\propto$$A^{-1/3}$ law of the well-known
breathing mode in medium-heavy and heavy nuclei.  Only the first mode
is accessible to AMD calculations, whereas the inclusion of the width
degree of freedom appears essential to describe the second one.  More
complicated spectral functions can be observed in our calculations
depending on the cluster structure of the nucleus.  The presented
calculations do not contain two-body collisions nor quantum branching,
meaning that they cannot describe the decay of the observed collective
motions, which are expected to considerably increase the fragmentation
of the monopole strength.  However, the dominant modes should be
preserved by such improved calculations \cite{colo92}.

Molecular dynamics calculation such as AMD and FMD therefore will be a
powerful tool to study the interplay among clustering structures,
collective resonances, and multiple breakup of various nuclei in
the nuclear chart including unstable nuclei.

\begin{acknowledgments}
This work is supported by the ANR project NExEN (ANR-07-BLAN-0256-02)
and by Grant-in-Aid for Scientific Research (KAKENHI) 20$\cdot$08814
and 21540253 from Japan Society for the Promotion of Science (JSPS).
T.F. acknowledges the support from ENSICAEN as post-doctoral
fellowship. K. H. O. H also thanks JSPS for his JSPS Postdoctoral
Fellowship for Foreign Researchers.
\end{acknowledgments}

\end{document}